\documentclass[%
 reprint,
 amsmath,amssymb,
 aps,
prl,
]{revtex4-2}

\usepackage[bookmarks,colorlinks]{hyperref}

\usepackage{graphicx}
\usepackage{dcolumn}
\usepackage{bm}

\usepackage[usenames,dvipsnames]{xcolor}

\begin{document}

\title{Optimal blind focusing on perturbation-inducing targets in sub-unitary complex media}

\author{Jérôme Sol}%
\author{Luc Le Magoarou}
\author{Philipp del Hougne}%
 \email{philipp.del-hougne@univ-rennes.fr}
\affiliation{Univ Rennes, INSA Rennes, CNRS, IETR - UMR 6164, F-35000 Rennes, France}%

\begin{abstract}

The scattering of waves in a complex medium is perturbed by polarizability changes or motion of embedded targets. These perturbations could serve as perfectly non-invasive guidestars for focusing on the targets. In this Letter, we theoretically derive a fundamental difference between these two perturbation types (the change of the scattering matrix is of rank one [two] for target polarizability changes [motion]) and identify accordingly optimal strategies to perfectly focus on the target in both cases. For target motion, at least two displacements are necessary. Furthermore, for the case of dynamic complex media additionally featuring parasitic perturbers, we establish a non-invasive scheme to achieve optimal time-averaged power delivery to a perturbation-inducing target. In all cases, no assumptions about the unitarity of the system's scattering matrix or the size of the perturbation are necessary. We experimentally demonstrate all results in the microwave regime using a strongly sub-unitary lossy chaotic cavity as complex medium. Our experiments highlight that the target's ``structural scattering'' is irrelevant [must be negligible] in the case of target polarizability changes [motion]. We expect our results to find applications in communications, cybersecurity, bioelectronics, flow-cytometry and self-propelled nano-swimmers.
\end{abstract}

\maketitle

Coherently focusing waves on a target embedded \textit{inside} a complex medium is a wave control primitive underpinning applications spanning from communications via sensing to wireless power transfer, across the entire spectrum of wave phenomena and scales~\cite{mosk2012controlling,gigan_rotter}. 
In this Letter, we define ``optimal focusing'' as delivering as much power as possible to the target.
Given the transmission coefficients $\mathbf{h}$ between the coherently controlled sources and the target, provably optimal focusing on the target by phase conjugation is possible~\cite{MaximumRatioTransmission,cheng2014focusing}. The difficulty usually lies in determining $\mathbf{h}$ given an unknown complex medium. 
Various guidestar-based focusing techniques were explored but they involve external invasive manipulation of the system to establish the guidestar~\cite{vellekoop2008demixing,hsieh2010digital,vellekoop2012digital,tao2012live,katz2014noninvasive,horstmeyer2015guidestar,del2017shaping}. 
Virtual guidestars still rely on exposing the system to external fields (distinct from the fields associated with the wave phenomenon intended to focus on the target) and sometimes suffer from limited resolution~\cite{larrat2010mr,xu2011time,judkewitz2013speckle,chaigne2014controlling,conkey2015super,ruan2017focusing}.
The holy grail would be to use a natural perturbation of the system originating from the target itself (e.g., a change of the target's polarizability or target motion) as a perfectly non-invasive guidestar.
The crux lies in retrieving $\mathbf{h}$ (or a vector $\mathbf{h^\prime}=c\mathbf{h}$ that is collinear with $\mathbf{h}$, where $c$ is an arbitrary complex-valued scalar) only based on how the system scatters known wavefronts for various perturbation states of the target.

Three techniques for focusing on perturbation-inducing targets inside complex media are reported in the literature. Let $\mathbf{S}$ be the system's scattering matrix that relates any incoming wavefront $\mathbf{x}$ and outgoing wavefront $\mathbf{y}$: $\mathbf{y} = \mathbf{S}\mathbf{x}$; the target-induced perturbation alters $\mathbf{S}$: $\mathbf{S} \rightarrow \mathbf{S} + \Delta \mathbf{S}$. First, Ref.~\cite{abboud2013noniterative} proposed to focus on a fault in a cable network using the first singular vector of $\Delta\mathbf{S}$~\footnote{Ref.~\cite{abboud2013noniterative} built on the DORT method for focusing on a \textit{static} and \textit{strongly} reflecting target in a \textit{weakly} scattering medium~\cite{prada1994eigenmodes,prada1996decomposition,popoff2011exploiting}.}; the appearance of the fault can be understood as a target polarizability change. Second, Refs.~\cite{ma2014time,zhou2014focusing} proposed the TRACK method that considers the change of the scattered field due to the target perturbation while the probing field remains the same: $\mathbf{h}^\prime_\mathrm{TRACK} = \Delta\mathbf{S}\mathbf{x}$. Third, Refs.~\cite{ambichl2017focusing,horodynski2020optimal} define $\mathbf{h^\prime}_\mathrm{GWS}$ as the conjugate of the first eigenvector of the generalized Wigner-Smith (GWS) operator $\mathbf{Q}_q = -\jmath \mathbf{S}^{-1} (\partial \mathbf{S} / \partial q)$ that is evaluated based on the derivative of $\mathbf{S}$ with respect to the perturbed parameter $q$. TRACK and GWS are agnostic to the perturbation's nature (polarizability change vs motion). 
In all cases, some experimental evidence of focusing was presented but without a claim or proof of having achieved \textit{optimal} focusing. 
Based on our results presented in this Letter, it is \textit{impossible} that the focusing in Refs.~\cite{ma2014time,zhou2014focusing,ambichl2017focusing,horodynski2020optimal} was close to optimal for the motion-induced perturbations.
Moreover, the GWS approach has so far not been analyzed based on a system model, and its assumptions that the perturbation strength is infinitesimal and that $\mathbf{S}$ is unitarity limit its scope of applicability; in most imaginable application scenarios, $\mathbf{S}$ is strongly sub-unitary due to significant absorption and/or leakage. A rigorous theoretical understanding of scattering in \textit{sub-unitary} unknown complex media involving perturbation-inducing targets and a technique for \textit{optimal} focusing on such targets is to date missing.

In this Letter, we fill this gap. Based on a recent model of tunable complex media~\cite{faqiri2022physfad,sol2023experimentally,del2023ris}, we show that there is a qualitative difference between various types of target-induced perturbations. 
If the target alters its polarizability, $\Delta\mathbf{S}$ is of rank one and optimal focusing simply requires a singular value decomposition (SVD) of $\Delta\mathbf{S}$. 
In the case of target motion, $\Delta\mathbf{S}$ is of rank two and we show that based on the SVDs of \textit{at least two} realizations of $\Delta\mathbf{S}$ originating from at least two target displacements, optimal focusing is possible.
Moreover, we show that the SVD approach can be extended to optimal non-invasive time-averaged power delivery to targets in \textit{dynamic} complex media where $\Delta\mathbf{S}$ contains contributions from parasitic perturbations and the perturbation-inducing target.
In all cases, no assumption about unitarity or the perturbation size is required. 
We corroborate all findings experimentally in the microwave domain using a chaotic cavity as complex medium. 
Along the way, we clarify the important role of the target's ``structural scattering cross-section''.

In the following, the superscripts $^T$, $^\star$ and $^\dagger$ and  denote the transpose, the conjugate and the transpose conjugate, respectively; $\breve{\mathbf{a}} = \mathbf{a} / \lVert \mathbf{a} \rVert_2$; $\left[ \mathbf{A} \right]_\mathcal{BC}$ denotes the block of the matrix $\mathbf{A}$ comprising rows [columns] whose indices are in the set $\mathcal{B}$ [$\mathcal{C}$].

To start, we briefly summarize the key features of our system model for tunable complex media that was recently validated experimentally in the context of ``smart'' radio environments~\cite{sol2023experimentally}. Applied to the problem of interest in this Letter, both the $N_\mathrm{A}$ antennas and the target (itself an antenna so that we can measure the ground-truth transmission coefficients) are modeled as point-like dipoles. 
In our microwave experiments, an electromagnetic antenna is a device that couples guided waves (incident via the single-mode coaxial cable that connects the antenna to a measurement device) to waves propagating in free space, and vice versa. The calibration plane is typically at the connection between the coaxial cable and the antenna such that the cable does not impact the measurement outcome. The ``dipole'' we are concerned with arises due to the charge separation between the two conductors of the coaxial cable in the calibration plane, implying that it is electrically very small (hence approximately point-like). To be clear, this definition is independent of the antenna design and we are hence \textit{not} referring to the entire antenna as the ``dipole''. 
An antenna usually scatters waves even when it is short-circuited (implying zero dipole moment). This so-called ``structural scattering''~\cite{king1949measurement,kahn1965minimum,hansen1989relationships} is significant even for electrically small antennas and plays a pivotal role below.

\begin{figure}
    \centering
    \includegraphics[width=\columnwidth]{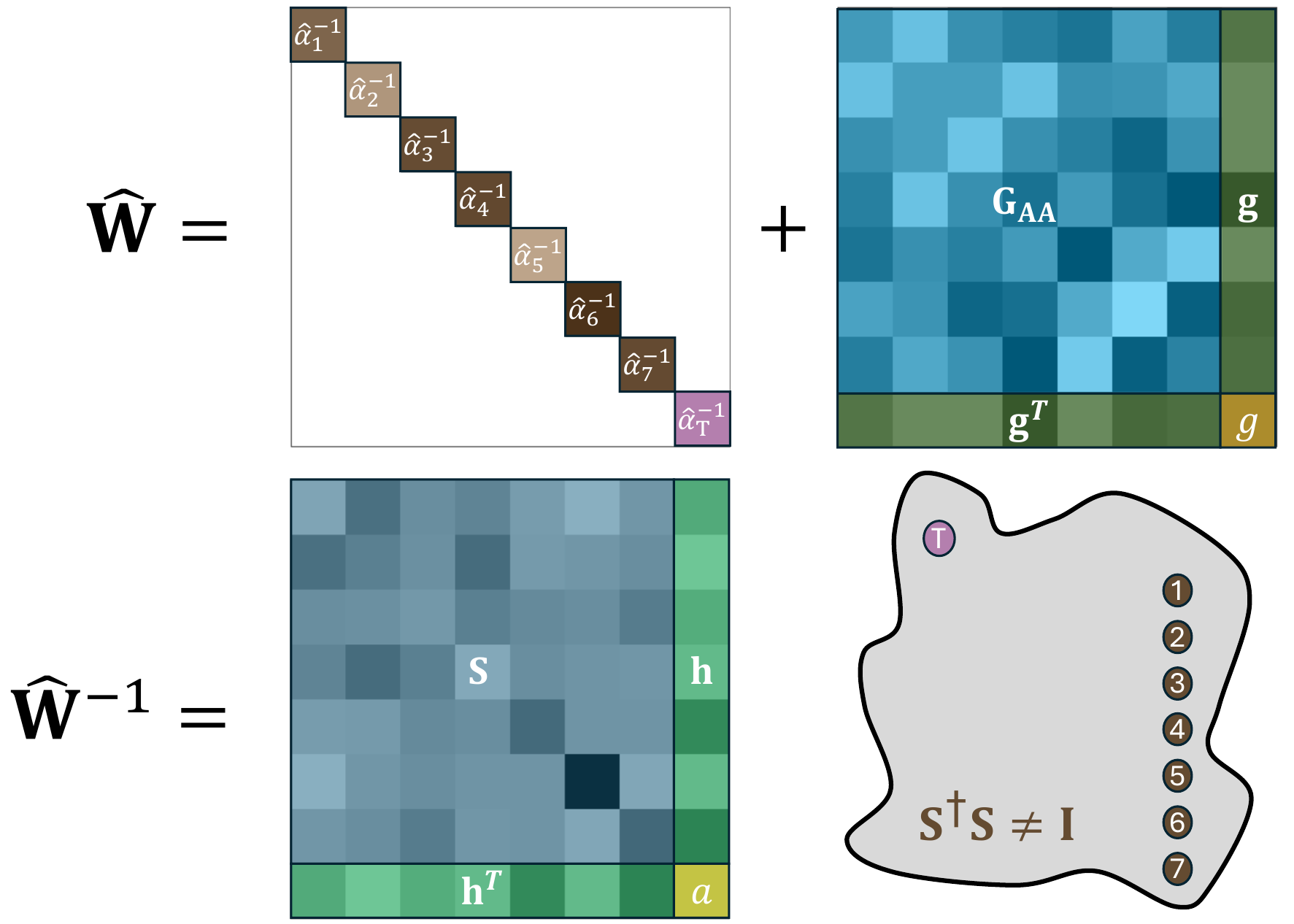}
    \caption{System model and sketch. }
    \label{Fig1}
\end{figure}

In our system model (illustrated in Fig.~\ref{Fig1})~\footnote{Our results on polarizability-changing targets could be derived equivalently (but more arduously) with a related model formulation in terms of impedances~\cite{tapie2023systematic} whereas this impedance formulation does not lend itself to the analysis of moving targets.}, the $i$th dipole is characterized by its polarizability $\alpha_i$ and coupled to the $j$th dipole via the \textit{background} Green's function $G_{ji}$ that takes into account the overwhelmingly complicated scattering in the unknown complex medium. 
In general, $G_{ii}\neq 0$ in a complex medium due to self-interactions.  
Let $\mathcal{A}$ and $\mathcal{T}$ be the sets of dipole indices representing the antennas and the target, respectively. We define an interaction matrix $\mathbf{W}$ whose $i$th diagonal entry is $\alpha_i^{-1} - G_{ii}$ while its $(i,j)$th off-diagonal entry is $-G_{ij}$. Our system model postulates that the scattering matrix defined by the antennas is proportional to the $\mathcal{AA}$ block of the inverse interaction matrix: $\mathbf{S} = \big[ \mathbf{\hat{W}}^{-1}\big]_\mathcal{AA}$, where the hat indicates that the quantity has absorbed multiplicative and additive factors that do not depend on the target polarizability (see Sec.~2.1.1 in Ref.~\cite{del2023ris} for details). 
In the following, we assume reciprocity ($\mathbf{S}=\mathbf{S}^T$, $\mathbf{\hat{W}}=\mathbf{\hat{W}}^T$) but \textit{not} unitarity ($\mathbf{S}\mathbf{S}^\dagger \neq \mathbf{I}_{N_\mathrm{A}}$). 
The sought-after transmission vector is $\mathbf{h}=\big[ \mathbf{\hat{W}}^{-1}\big]_\mathcal{AT}$. As mentioned earlier, knowing some vector $\mathbf{h^\prime}$ that is collinear with $\mathbf{h}$ is sufficient for our goal of blind optimal focusing on the target, where ``blind'' refers to the fact that the spatial coordinates of the target are unknown.

The first target-induced perturbation we examine is a change of its inverse polarizability: $\alpha_\mathrm{T}^{-1} \rightarrow \alpha_\mathrm{T}^{-1}+\Delta\alpha_\mathrm{T}^{-1}$. 
The inverse of the new interaction matrix $\mathbf{{\hat{W}}}_{\mathrm{p}}$ after the target's change of polarizability can be related to that of the previous interaction matrix $\mathbf{\hat{W}}$ via the Woodbury identity~\cite{hager1989updating,prod2023efficient}. Straightforward algebraic manipulations (see SM) yield
\begin{equation}
    \Delta\mathbf{S}_{\mathrm{p}} = \left[ \mathbf{{\hat{W}}}_{\mathrm{p}}^{-1}\right]_\mathcal{AA} - \left[ \mathbf{{\hat{W}}}^{-1}\right]_\mathcal{AA} = -k \mathbf{h}\mathbf{h}^T.
    \label{eq_DeltaS_pol}
\end{equation}
Given Eq.~(\ref{eq_DeltaS_pol}), it is clear that $\Delta\mathbf{S}_{\mathrm{p}}$ is a rank-one matrix whose first (and only) left singular vector is collinear with $\mathbf{h}$, implying the ability to achieve blind optimal focusing~\footnote{The importance of $\Delta\mathbf{S}$ being of rank one to achieve optimal focusing with the first left singular vector of $\Delta\mathbf{S}$ has also been recognized in Refs.~\cite{del2021experimental,yeo2022time}.}. 
$\Delta\mathbf{S}_{\mathrm{p}}=\mathbf{U}_{\mathrm{p}}\mathbf{\Sigma}_{\mathrm{p}}\mathbf{V}_{\mathrm{p}}^\dagger$ is the SVD, where $\mathbf{\Sigma}_{\mathrm{p}}$ is a diagonal matrix containing the singular values (in descending order); the $i$th column of $\mathbf{U}_{\mathrm{p}}$ (resp. $\mathbf{V}_{\mathrm{p}}$) is the $i$th left (resp. right) singular vector ${\mathbf{u}_{\mathrm{p}}}_i$ (resp. ${\mathbf{v}_{\mathrm{p}}}_i$) of $\Delta\mathbf{S}_{\mathrm{p}}$. The provably optimal ``SVD approach'' yields $\mathbf{h}_{\mathrm{p}}^\prime={\mathbf{u}_{\mathrm{p}}}_1$. 
Our derivation did not make any assumptions regarding the unitarity of $\mathbf{S}$ or the magnitude of the polarizability change.
TRACK is equivalent to right-multiplying $\Delta\mathbf{S}_{\mathrm{p}}$ with some non-zero wavefront $\mathbf{x}$; GWS left-multiplies $\Delta\mathbf{S}_{\mathrm{p}}$ (taken as approximation of $\partial \mathbf{S}/\partial\alpha_\mathrm{T}$) with $-\jmath \mathbf{S}^{-1}$ and extracts the first (and only) eigenvector. Hence, TRACK and GWS in principle also identify vectors that are collinear with $\mathbf{h}$ (see SM), enabling optimal focusing. 
The assumptions in Refs.~\cite{ambichl2017focusing,horodynski2020optimal} that $\mathbf{S}^\dagger\mathbf{S}=\mathbf{I}_{N_\mathrm{A}}$ and $\Delta\alpha_\mathrm{T}^{-1}\rightarrow 0$ are unnecessary.

In our microwave experiments, a change of polarizability takes the form of changing the termination of the target antenna's port: instead of connecting it to the coaxial cable, it can be, for instance, short-circuited or open-circuited. 
Antennas that self-modulate their load impedance arise in many practical contexts: ``reconfigurable intelligent surfaces'' in next-generation wireless networks~\cite{subrt2012intelligent,Liaskos_Visionary_2018,del2023ris}, backscatter communication schemes~\cite{liu2013ambient,zhao2020metasurface}, spy equipment~\cite{brooker2013lev}, biomedical ingestible devices~\cite{mandal2008power}. While Ref.~\cite{abboud2013noniterative} performed SVD-based focusing without discussing its optimality, Refs.~\cite{del2021coherent,yeo2022time} recently observed the optimality of GWS-based focusing on load-impedance-modulated antennas in chaotic cavities without explaining how that was possible despite obvious violations of fundamental GWS assumptions (sub-unitary system, large change in load impedance). This is now clear. 
Two additional questions remain: (i) Does the target antenna design play any role? (Ref.~\cite{yeo2022time} assumes it should be electrically small.) (ii) Are the three approaches (TRACK, GWS, SVD) fully equivalent in the presence of noise?

\begin{figure}
    \centering
    \includegraphics[width=\columnwidth]{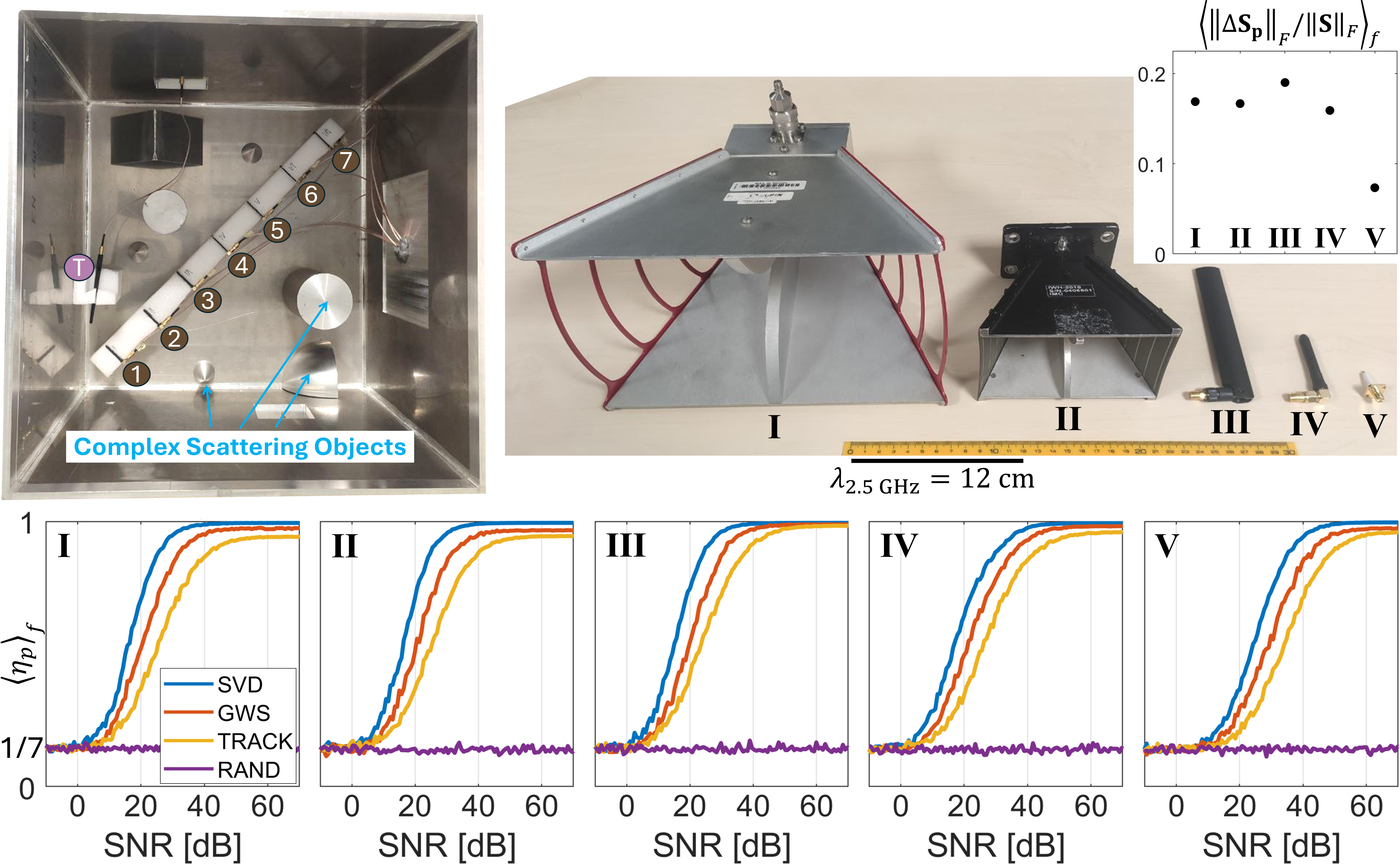}
    \caption{Blind optimal focusing on polarizability-changing target (load-impedance-modulated target antenna) inside unknown chaotic microwave cavity for five target antenna types. }
    \label{Fig2}
\end{figure}

We conducted experiments with five distinct target antennas of drastically different sizes in the chaotic microwave cavity seen in Fig.~\ref{Fig2} ($59\times 60 \times 58 \mathrm{cm}^3$; composite quality factor: $Q=758$; modal overlap: $\mathcal{N}=4$). $N_\mathrm{A}=7$ antennas were coupled to the system. Our results in Fig.~\ref{Fig2} show that for high signal-to-noise ratios (SNRs), the antenna design does $\textit{not}$ play any role: $\eta_p = \big|\mathbf{h}^\dagger {\mathbf{h}_{\mathrm{p}}^\prime} \big|^2/ \big|\mathbf{h}^\dagger \mathbf{\breve{h}} \big|^2 = 100\%$ in all cases. This observation makes sense because the target antenna's size impacts its structural scattering cross-section but this is fixed and an indistinguishable part of the background scattering. 
However, $\lVert \Delta\mathbf{S}_{\mathrm{p}} \rVert_F$ depends on the target antenna type because the latter impacts the coupling strength (background Green's functions) between the target dipole and the $N_\mathrm{A}$ source dipoles. Hence, for lower SNRs the performance degrades more rapidly in case V than in cases I-IV.
We also observe that TRACK and GWS fall slight short of the SVD performance. This also makes sense because of their avoidable vulnerabilities (see SM), and additionally TRACK is fundamentally disadvantaged in this comparison because it relies on $N_\mathrm{A}$ fewer measurements.
Below SNRs of roughly 0~dB, all methods fail (their performance equals the baseline of taking a normalized random vector as $\mathbf{h^\prime}$, yielding $\langle\eta_p\rangle_f = 1/N_\mathrm{A} = 1/7$).

\begin{figure}
    \centering
    \includegraphics[width=\columnwidth]{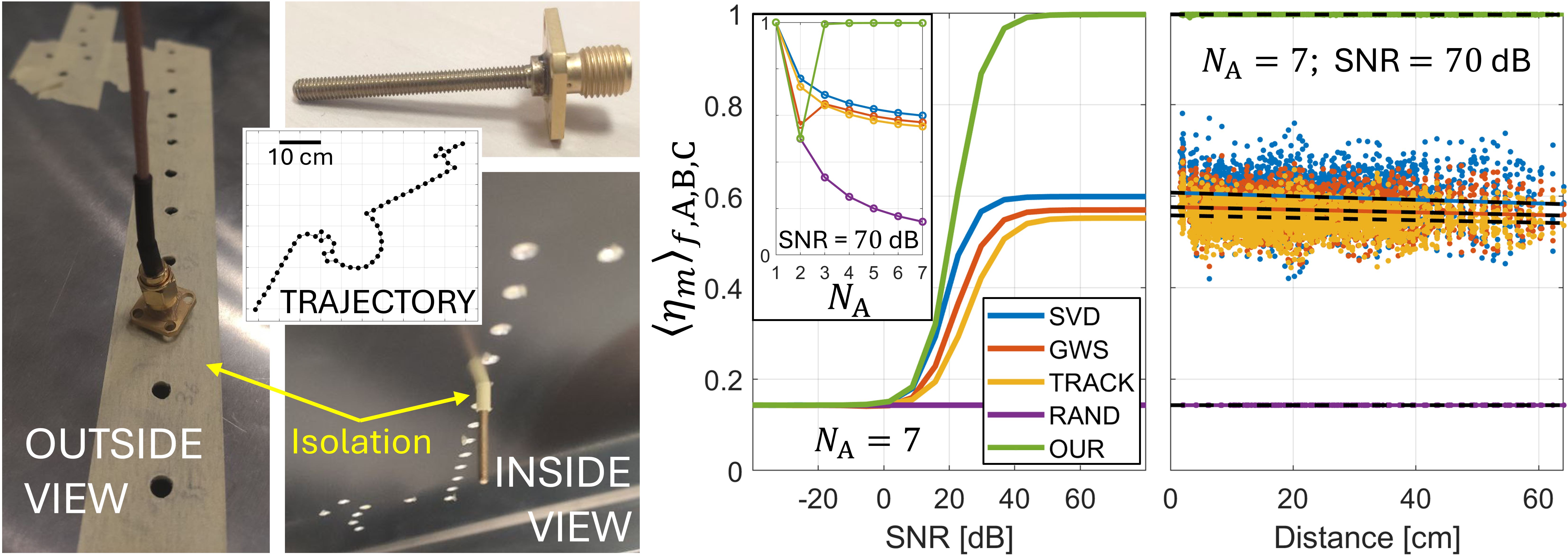}
    \caption{Blind optimal focusing on moving target inside unknown chaotic microwave cavity. }
    \label{Fig3}
\end{figure}

We now proceed to examine a second (more challenging) target-induced perturbation: motion. Whereas TRACK and GWS do not distinguish between different types of perturbations, our model-based analysis reveals an important difference compared to the previously examined change of target polarizability. The interaction matrix $\mathbf{{\hat{W}}}_{\mathrm{m}}$ after the target motion and the previous interaction matrix $\mathbf{\hat{W}}$ differ regarding an entire column and an entire row because all background Green's functions involving the target (denoted by $\mathbf{g}$, $\mathbf{g}^T$ and $g$ in Fig.~\ref{Fig1}) change.
Hence, $\Delta\mathbf{S}_{\mathrm{m}}$ (change of $\mathbf{S}$ due to target motion) must be of rank two whereas $\Delta\mathbf{S}_{\mathrm{p}}$ (change of $\mathbf{S}$ due to target polarizability change) was of rank one. 
The Woodbury identity yields an exact expression for $\Delta\mathbf{S}_{\mathrm{m}}$ (see Ref.~\cite{prod2023efficient} and SM for details):
\begin{equation}
\begin{split}
      \Delta\mathbf{S}_{\mathrm{m}} = \left( K_{11}+K_{12}d_\mathrm{T}\right)\mathbf{e}\mathbf{h}^T \\ + \left(K_{21}+K_{22}d_\mathrm{T}\right)\mathbf{h} \mathbf{h}^T + K_{12}\mathbf{e}\mathbf{e}^T +K_{22}\mathbf{h} \mathbf{e}^T,
\end{split}
\label{eq_DSm}
\end{equation}
where $\mathbf{e} = \mathbf{S}\mathbf{d}$, $\mathbf{d} = \big[ \mathbf{\hat{W}}_{\mathrm{m}}\big]_\mathcal{AT} - \big[ \mathbf{\hat{W}}\big]_\mathcal{AT}$,  and $d_\mathrm{T}$, $K_{11}$, $K_{12}$, $K_{21}$, and $K_{22}$ are complex-valued scalars. Inspection of Eq.~(\ref{eq_DSm}) reveals that the first two left singular vectors ${\mathbf{u}_{\mathrm{m}}}_1$ and ${\mathbf{u}_{\mathrm{m}}}_2$ of $\Delta\mathbf{S}_{\mathrm{m}}$ span a space containing $\mathbf{h}$ and $\mathbf{e}$. Hence, ${\mathbf{u}_{\mathrm{m}}}_1$ is a linear combination of $\mathbf{h}$ and $\mathbf{e}$ that is, in general, \textit{not} collinear with $\mathbf{h}$.
Consequently, \textit{perfect} focusing on a moving target based on $\Delta\mathbf{S}_{\mathrm{m}}$ is impossible. As seen in Fig.~\ref{Fig3}, TRACK, GWS or SVD still achieve some focusing on average but fall significantly short of the optimum.

Fortunately, blind optimal focusing on a moving target is nonetheless possible. To that end, note that every target displacement yields a distinct $\mathbf{e}$. Hence, given at least three measurements $\mathbf{S}_{\mathrm{A}}$, $\mathbf{S}_{\mathrm{B}}$ and $\mathbf{S}_{\mathrm{C}}$ corresponding to three arbitrary distinct positions A, B and C of the target, we evaluate the two rank-two matrices $\Delta\mathbf{S}_{\mathrm{m}}^\mathrm{AC} = \mathbf{S}_{\mathrm{C}}-\mathbf{S}_{\mathrm{A}}$ and $\Delta\mathbf{S}_{\mathrm{m}}^\mathrm{BC}= \mathbf{S}_{\mathrm{C}}-\mathbf{S}_{\mathrm{B}}$ (assuming, without loss of generality, that we wish to focus on the target while it is at position C). Then, we extract $\mathbf{h}_{\mathrm{m}}^\prime$ as the intersection of the spaces spanned by the two first singular vectors of $\Delta\mathbf{S}_{\mathrm{m}}^\mathrm{AC}$ and $\Delta\mathbf{S}_{\mathrm{m}}^\mathrm{BC}$. To find $\mathbf{h}_{\mathrm{m}}^\prime$, we define the matrix $\mathbf{H} = [\mathbf{u}_1^\mathrm{AC},\mathbf{u}_2^\mathrm{AC},\mathbf{u}_1^\mathrm{BC},\mathbf{u}_2^\mathrm{BC}]$ and define $\mathbf{h}_{\mathrm{m}}^\prime$ as the first left singular vector of $\mathbf{H}$. 
$\mathbf{h}_{\mathrm{m}}^\prime$ is collinear with $\mathbf{h}_{\mathrm{C}}$ and hence optimal for focusing on the target while it is at position C. Our method requires $N_\mathrm{A}\geq 3$ to ensure that the SVD of $\mathbf{H}$ can separate its three constituent vectors (see SM).

To experimentally demonstrate blind \textit{optimal} focusing on a moving target, the target's structural scattering must be minimal; otherwise, it would be as if parts of the system move with the target dipole such that based on realizations of $\Delta\mathbf{S}_{\mathrm{m}}$ the contribution from the target dipole could not be isolated. We heuristically found that the antenna shown in Fig.~\ref{Fig3} has minimal structural scattering. Moving this target along the indicated trajectory, we evaluated the efficiency $\eta_m = \big|\mathbf{h}_\mathrm{C}^\dagger \mathbf{h}_{\mathrm{m}}^\prime\big|^2 / \big| \mathbf{h}_\mathrm{C}^\dagger \breve{\mathbf{h}}_\mathrm{C} \big|^2 $  of focusing on C for all possible choices of A and B and all frequency points. Our method achieves perfect focusing for SNRs above 40~dB at all considered frequency points, irrespective of the relative distances between A, B and C. In contrast, the benchmarks SVD, GWS and TRACK only reach around $60\%$ efficiency on average and slightly deteriorate as the distance between A and C increases.

So far, we only considered complex media that are perfectly static except for the target-induced perturbation. Practical applications often involve \textit{dynamic} complex media~\cite{del2020robust,saigre2023self,mididoddi2023threading} in which parasitic perturbations not originating from the target distort $\Delta\mathbf{S}$ and thereby the discussed estimation of $\mathbf{h}^\prime$. 
Specifically, we now discuss parasitic perturbations occurring on time scales faster than that at which the previously outlined techniques can be performed -- otherwise the latter could simply be applied repeatedly in a loop. Hence, the goal is to find a wavefront $\mathbf{\mathring{x}}$ that delivers the largest possible power to the target \textit{on average} (over realizations of the parasitic perturbations). The time-averaged delivered power is $\left| \langle \mathbf{h}^T(t) \mathbf{\mathring{x}}\rangle_t \right|^2 = \left| \langle \mathbf{h}^T(t) \rangle_t \mathbf{\mathring{x}} \right|^2 $, revealing that the optimal choice of $\mathbf{\mathring{x}}$ is collinear with $\overline{\mathbf{h}}^* = \langle\mathbf{h}(t)\rangle_t^*$. 
However, $\overline{\mathbf{h}}$ cannot be extracted directly from $\Delta\mathbf{S}$: for a target polarizability change, any given $\Delta\mathbf{S}_{\mathrm{p}}(t) = \mathbf{S}(t) - \mathbf{S}(t-\Delta t)$ has $\mathrm{min}(1+m,N_\mathrm{A})$ non-zero singular values if $m \leq N_\mathrm{S}$ of the $N_\mathrm{S}$ parasitic perturbers (also described as dipoles in our system model) changed their polarizabilities between time instants $t-\Delta t$ and $t$. 
The left singular vectors of $\Delta\mathbf{S}_{\mathrm{p}}(t)$ are hence linear combinations of $\mathbf{h}(t)$ and the $m$ transmission vectors $\mathbf{l}_v(t)$ to the $m$ parasitic perturbers that were active in the considered time interval (see SM). 
$\mathbf{h}$ and $\mathbf{l}_v$ are composed of a static and a dynamic component: $\mathbf{h}(t) = \overline{\mathbf{h}} + \mathbf{\check{h}}(t)$ and $\mathbf{l}_v(t) = \overline{\mathbf{l}}_v + \mathbf{\check{l}}_v(t)$.

Let us assume that the target periodically changes its polarizability at a known rate $1/\Delta t$ that is not correlated with the perturbers. Then, to extract $\overline{\mathbf{h}}$, we stack in a matrix $\mathbf{F}$ the first $w$ left singular vectors of $\gamma$ realizations of $\Delta\mathbf{S}_{\mathrm{p}}(t)$ measured at time instants $t$ separated by $\Delta t$. 
We define our estimate $\overline{\mathbf{h}}^\prime$ as the first left singular vector of $\mathbf{F}$. 
If  $N_\mathrm{A} \geq 2+N_\mathrm{S}$ so that we can choose $w \geq 1+N_\mathrm{S}$ and
$\gamma \rightarrow \infty$, we expect that $\overline{\mathbf{h}}^\prime$ is collinear with $\overline{\mathbf{h}}$. To understand why, it is sufficient to note that $\overline{\mathbf{h}}$ is the most present vector in $\mathbf{F}$: every realization of $\mathbf{\check{h}}(t)$ and $\mathbf{\check{l}}_v(t)$ is uncorrelated with the others, and any given $\overline{\mathbf{l}}_v$ appears statistically only in every second realization of $\Delta\mathbf{S}_{\mathrm{p}}$. If the sources-target coupling is stronger than the sources-perturber coupling, smaller values of $w$ can be chosen since $\mathbf{h}$ will be present in the first few left singular vectors of $\Delta\mathbf{S}_{\mathrm{p}}(t)$, and fewer realizations are necessary for $\overline{\mathbf{h}}^\prime$ to converge toward being collinear with $\overline{\mathbf{h}}$. Hence the target antenna type is expected to matter.

\begin{figure}
\centering
    \includegraphics[width=\columnwidth]{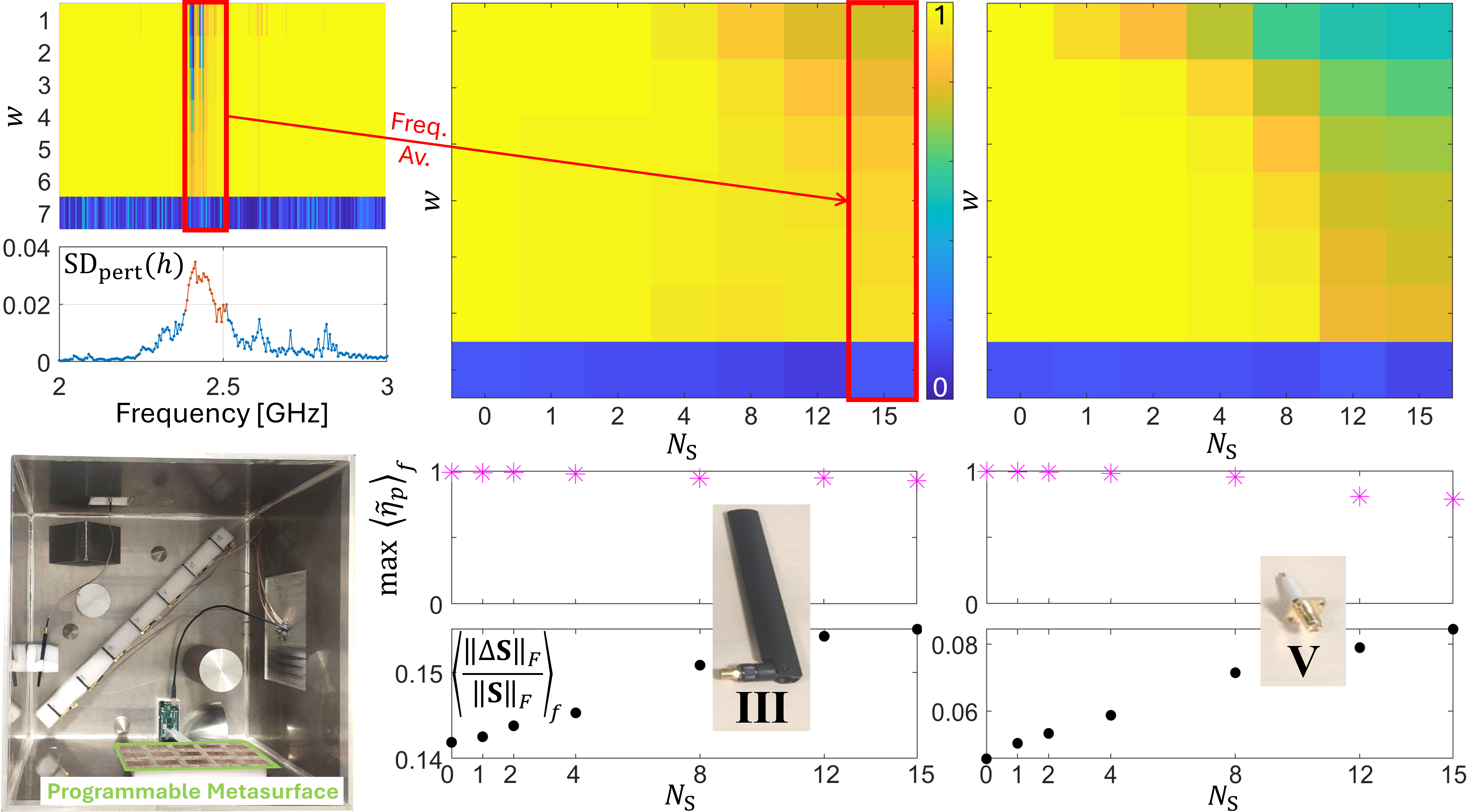}
    \caption{Blind non-invasive optimal time-averaged power delivery to polarizability-changing target in \textit{dynamic} complex environment emulated by $N_\mathrm{S}\leq 15$ 1-bit programmable meta-atoms working around 2.45~GHz. All results are for $\gamma=10^3$ and $\mathrm{SNR}=50\ \mathrm{dB}$.}
    \label{Fig4}
\end{figure}

To systematically explore our approach, we emulate the parasitic perturbations with a 15-element programmable metasurface (PM, see Fig.~\ref{Fig4}). Within the PM's operating band ($2.39 - 2.51\ \mathrm{GHz}$), each of its meta-atoms can be modelled as a dipole whose polarizability can be individually switched between two values~\cite{sol2023experimentally}. By sweeping $N_\mathrm{S}\leq15$ meta-atoms through random configurations while keeping the remaining ones static, a dynamic environment with a given number of (identical) parasitic perturbers can be emulated~\cite{ahmed2023over}.

For target antenna type III, $\lVert \Delta\mathbf{S}_{\mathrm{p}} \rVert_F$ only weakly depends on $N_\mathrm{S}$ such that $\mathbf{h}(t)$ is present in the first few left singular vectors of $\Delta\mathbf{S}_{\mathrm{p}}(t)$. Hence, even though $N_\mathrm{A}=7$ forces us to use $w<1+N_\mathrm{S}$ for $N_\mathrm{S}>5$, we achieve $\langle\tilde{\eta}_p\rangle_f = \langle \big| \overline{\mathbf{h}}^\dagger \overline{\mathbf{h}}^\prime \big| ^2/ \big| \overline{\mathbf{h}}^\dagger \breve{\overline{\mathbf{h}}} \big|^2 \rangle_f = 93\%$ for $N_\mathrm{S}=15$ (averaging only across $2.39 - 2.51\ \mathrm{GHz}$). 
For target antenna type V, $\lVert \Delta\mathbf{S}_{\mathrm{p}} \rVert_F$ significantly depends on $N_\mathrm{S}$; for larger values of $N_\mathrm{S}$, $\mathbf{h}(t)$ is thus not guaranteed to be entirely present in the first few left singular vectors of $\Delta\mathbf{S}_{\mathrm{p}}$ that are stacked in $\mathbf{F}$. Hence, $\overline{\mathbf{h}}^\prime$ is not perfectly collinear with $\overline{\mathbf{h}}$. Nonetheless, we achieve $\langle\tilde{\eta}_p\rangle_f =79\%$ for $N_\mathrm{S}=15$.
Overall, the proposed method is seen to be very effective at approaching optimal time-averaged power delivery to a polarization-changing target in a \textit{dynamic} complex medium.

Finally, we remark that in case only an \textit{off-diagonal} block of $\mathbf{S}$ can be measured, optimal focusing is still possible with the techniques outlined in this Letter except that instead of the left singular vector(s) one should use the conjugate(s) of the right singular vector(s). Theoretical details and experimental validation are provided in the SM. Such transmission-only scenarios arise in simplex wireless communication systems and optical wavefront shaping experiments.

To summarize, our model-based analysis revealed a fundamental difference between perturbations of complex media due to polarizability changes vs motion of an \textit{embedded} target, allowing us to theoretically and experimentally achieve non-invasive blind optimal focusing on the target in both cases. We also experimentally demonstrated non-invasive blind optimal time-averaged power delivery to a target \textit{inside} a dynamic complex medium. Fundamental assumptions of the GWS approach (unitarity, infinitesimal change) are unnecessary for optimal focusing.

Looking forward, we expect our techniques for focusing on polarizability-changing targets to find practical microwave applications in communications, cybersecurity and bioelectronics. Meanwhile, we expect our technique for focusing on moving targets to find applications in optics where targets of interest include gold nanobeads~\cite{popoff2011exploiting}, self-propelled nano-swimmers converting received energy into motion~\cite{bechinger2016active}, and cells in flow-cytometry~\cite{zhou2014focusing}.


\providecommand{\noopsort}[1]{}\providecommand{\singleletter}[1]{#1}%

\clearpage

\onecolumngrid

\renewcommand{\thesection}{S\Roman{section}}
\setcounter{section}{0}
\renewcommand{\thefigure}{S\arabic{figure}}
\setcounter{figure}{0}
\renewcommand{\theequation}{S\arabic{equation}}
\setcounter{equation}{0}

\centerline{\bf SUPPLEMENTAL MATERIAL:  }
\centerline{\bf \large Optimal blind focusing on perturbation-inducing targets in sub-unitary complex media  }
\bigskip

\centerline{Jérôme Sol, Luc Le Magoarou, and Philipp del Hougne\textsuperscript{*}}

\bigskip

\centerline{\small{\textit{Univ Rennes, INSA Rennes, CNRS, IETR - UMR 6164, F-35000 Rennes, France}}}
\centerline{\textsuperscript{*}\small{Correspondence to \href{mailto:philipp.del-hougne@univ-rennes.fr}{philipp.del-hougne@univ-rennes.fr}}}

\tableofcontents

\clearpage
\section{Derivation of Eq.~(1)}
\label{SupSecI}

After a target polarizability change, the updated inverse interaction matrix $\mathbf{{\hat{W}}_\mathrm{p}}^{-1}$ is related to the previous inverse interaction matrix $\mathbf{{\hat{W}}}^{-1}$ according to the Woodbury identity~\cite{hager1989updating,prod2023efficient} via the following relation:
\begin{equation}
    \mathbf{{\hat{W}}_\mathrm{p}}^{-1} = \left( \mathbf{{\hat{W}}} + \mathbf{UCV} \right)^{-1} = \mathbf{{\hat{W}}}^{-1} - \mathbf{{\hat{W}}}^{-1} \mathbf{U} \left(\mathbf{C}^{-1} + \mathbf{V} \mathbf{{\hat{W}}}^{-1} \mathbf{U} \right)^{-1} \mathbf{V} \mathbf{{\hat{W}}}^{-1}.
    \label{eq_woodbury}
\end{equation}
\noindent Herein,
\begin{subequations}
    \begin{equation}
        \mathbf{C} = \Delta \alpha_\mathrm{T}^{-1},
    \end{equation}
    \begin{equation}
        \mathbf{U} = \left[ \mathbf{0}_{N_\mathrm{A}} \ 1 \right]^T = \mathbf{V}^T,
    \end{equation}
\end{subequations}
where $\Delta \alpha_\mathrm{T}^{-1}$ is a complex-valued scalar and $\mathbf{0}_{N_\mathrm{A}}$ denotes a $N_\mathrm{A}$-element vector whose entries are zero. As pointed out in Ref.~\cite{prod2023efficient}, $\mathbf{U}$ and $\mathbf{V}$ act as ``selectors'' here, implying that
\begin{subequations}
    \begin{equation}
        \mathbf{{\hat{W}}}^{-1} \mathbf{U} = \begin{bmatrix} 
	\mathbf{h}  \\ 	a  \\ \end{bmatrix}.
    \end{equation}
    \begin{equation}
         \mathbf{V} \mathbf{{\hat{W}}}^{-1} \mathbf{U} = \left[ \mathbf{\hat{W}}^{-1}\right]_\mathcal{TT} = a.
    \end{equation}
    \begin{equation}
        \mathbf{V} \mathbf{{\hat{W}}}^{-1} = \begin{bmatrix} 
	\mathbf{h}^T  & 	a   \end{bmatrix}.
    \end{equation}
\end{subequations}
Substituting these results into Eq.~(\ref{eq_woodbury}), we find
\begin{equation}
        \mathbf{{\hat{W}}_\mathrm{p}}^{-1}  = \mathbf{{\hat{W}}}^{-1} - \begin{bmatrix} 
	\mathbf{h}  \\ 	a  \\ \end{bmatrix} \left( \frac{1}{\Delta\alpha_\mathrm{T}^{-1}} + a \right)^{-1} \begin{bmatrix} 
	\mathbf{h}^T  & 	a   \end{bmatrix}
\end{equation}
which, upon defining the complex-valued scalar $k = \left( (1/\Delta\alpha_\mathrm{T}^{-1}) + a \right)^{-1}$, yields:
\begin{equation}
  \mathbf{{\hat{W}}}_\mathrm{p}^{-1} = \mathbf{\hat{W}}^{-1} - \begin{bmatrix} 
	\mathbf{h}  \\ 	a  \\ \end{bmatrix} k  \begin{bmatrix} 
	\mathbf{h}^T  & 	a   \end{bmatrix}.
 \label{eq1main}
\end{equation}

Finally, we obtain Eq.~(1) from the main text: 
\begin{equation}
    \Delta\mathbf{S}_\mathrm{p} = \left[ \mathbf{{\hat{W}}}_\mathrm{p}^{-1}\right]_\mathcal{AA} - \left[ \mathbf{{\hat{W}}}^{-1}\right]_\mathcal{AA} = -k \mathbf{h}\mathbf{h}^T.
    \label{eq_DeltaS_pol}
\end{equation}

\clearpage
\section{Derivation of Eq.~(2)}

After a target displacement, the updated inverse interaction matrix $\mathbf{{\hat{W}}_\mathrm{m}}^{-1}$ is related to the previous inverse interaction matrix $\mathbf{{\hat{W}}}^{-1}$ according to the Woodbury identity~\cite{hager1989updating,prod2023efficient} via the following relation:
\begin{equation}
    \mathbf{{\hat{W}}_\mathrm{m}}^{-1} =  \left( \mathbf{{\hat{W}}} + \mathbf{A}\mathbf{I}_2\mathbf{B} \right)^{-1} = \mathbf{{\hat{W}}}^{-1} - \mathbf{{\hat{W}}}^{-1} \mathbf{A} \left(\mathbf{I}_2 + \mathbf{B} \mathbf{{\hat{W}}}^{-1} \mathbf{A} \right)^{-1} \mathbf{B} \mathbf{{\hat{W}}}^{-1}.
    \label{eq_woodbury2}
\end{equation}
Herein, $\mathbf{I}_2$ denotes the $2 \times 2$ identity matrix and
\begin{subequations}
    \begin{equation}
        \mathbf{A} = \begin{bmatrix}  \mathbf{d}  &  \mathbf{0}_{N_\mathrm{A}} \\ 0 & 1    \\ \end{bmatrix} ,
    \end{equation}
    \begin{equation}
        \mathbf{B} = \begin{bmatrix}    \mathbf{0}_{N_\mathrm{A}}^T & 1     \\ \mathbf{d}^T  & d_\mathrm{T} \\  \end{bmatrix} ,
    \end{equation}
\end{subequations}
where $\mathbf{0}_{N_\mathrm{A}}$ denotes a zero vector of dimensions $N_\mathrm{A} \times 1$, $\mathbf{d} = \left[ \mathbf{\hat{W}}_\mathrm{m}\right]_\mathcal{AT} - \left[ \mathbf{\hat{W}}\right]_\mathcal{AT}$ and $d_\mathrm{T} = \left[ \mathbf{\hat{W}}_\mathrm{m}\right]_\mathcal{TT} - \left[ \mathbf{\hat{W}}\right]_\mathcal{TT}$. Note that $d_\mathrm{T}$ represents the change of the target's self-interaction due to its displacement; meanwhile, the target's polarizability remains unchanged. To avoid that $d_\mathrm{T}$ is incorrectly taken into account twice, it only appears in the definition of $\mathbf{B}$. Indeed, we can verify that the definitions of $\mathbf{A}$ and $\mathbf{B}$ yield the correct form of $\mathbf{{\hat{W}}_\mathrm{m}} - \mathbf{{\hat{W}}}$:
\begin{equation}
    \mathbf{{\hat{W}}_\mathrm{m}} - \mathbf{{\hat{W}}} = \mathbf{A}\mathbf{I}_2\mathbf{B} = \begin{bmatrix}  \mathbf{d}  &  \mathbf{0}_{N_\mathrm{A}} \\ 0 & 1    \\ \end{bmatrix}  \begin{bmatrix}    \mathbf{0}_{N_\mathrm{A}}^T & 1     \\ \mathbf{d}^T  & d_\mathrm{T} \\  \end{bmatrix} =  \begin{bmatrix}    \mathbf{0}_{N_\mathrm{A}\times N_\mathrm{A}} & \mathbf{d}     \\ \mathbf{d}^T  & d_\mathrm{T} \\  \end{bmatrix},
\end{equation}
where $\mathbf{0}_{N_\mathrm{A}\times N_\mathrm{A}}$ denotes a zero matrix of dimensions $N_\mathrm{A} \times N_\mathrm{A}$.

We begin by evaluating the following terms that appear in Eq.~(\ref{eq_woodbury2}):
\begin{subequations}
    \begin{equation}
        \mathbf{{\hat{W}}}^{-1} \mathbf{A} = \begin{bmatrix}  \mathbf{S}  &  \mathbf{h} \\ \mathbf{h}^T & a    \\ \end{bmatrix} \begin{bmatrix}  \mathbf{d}  &  \mathbf{0}_{N_\mathrm{A}} \\ 0 & 1    \\ \end{bmatrix} = \begin{bmatrix}  \mathbf{S}\mathbf{d}  &  \mathbf{h} \\ \mathbf{h}^T\mathbf{d} & a    \\ \end{bmatrix}.
    \end{equation}
        \begin{equation}
        \mathbf{B}\mathbf{{\hat{W}}}^{-1}  = \begin{bmatrix}    \mathbf{0}_{N_\mathrm{A}}^T & 1     \\ \mathbf{d}^T  & d_\mathrm{T} \\  \end{bmatrix} \begin{bmatrix}  \mathbf{S}  &  \mathbf{h} \\ \mathbf{h}^T & a    \\ \end{bmatrix}  = \begin{bmatrix} \mathbf{h}^T  &  a \\ \mathbf{d}^T\mathbf{S} + d_\mathrm{T}\mathbf{h}^T & \mathbf{d}^T\mathbf{h}+d_\mathrm{T} a    \\ \end{bmatrix}.
    \end{equation}
\end{subequations}
Moreover, we define
\begin{equation}
    \mathbf{K} = \begin{bmatrix} K_{11}  &  K_{12} \\ K_{21} & K_{22}    \\ \end{bmatrix} = \left(\mathbf{I}_2 + \mathbf{B} \mathbf{{\hat{W}}}^{-1} \mathbf{A} \right)^{-1}. 
\end{equation}
For the purposes of our present analysis, there is no need to work out an exact expressions for $\mathbf{K}$ in terms of $\mathbf{S}$, $\mathbf{h}$, $a$, $\mathbf{d}$ and $d_\mathrm{T}$.
Next, we evaluate the second term on the RHS of Eq.~(\ref{eq_woodbury2}):
\begin{equation}
\begin{split}
    &\mathbf{{\hat{W}}}^{-1} \mathbf{A} \mathbf{K} \mathbf{B} \mathbf{{\hat{W}}}^{-1} = \begin{bmatrix}  \mathbf{S}\mathbf{d}  &  \mathbf{h} \\ \mathbf{h}^T\mathbf{d} & a    \\ \end{bmatrix} \begin{bmatrix} K_{11}  &  K_{12} \\ K_{21} & K_{22}    \\ \end{bmatrix} \begin{bmatrix} \mathbf{h}^T  &  a \\ \mathbf{d}^T\mathbf{S} + d_\mathrm{T}\mathbf{h}^T & \mathbf{d}^T\mathbf{h}+d_\mathrm{T} a    \\ \end{bmatrix} 
    \\&=  \begin{bmatrix} K_{11}\mathbf{S}\mathbf{d}+K_{21}\mathbf{h}  &  K_{12}\mathbf{S}\mathbf{d} + K_{22}\mathbf{h} \\ K_{11}\mathbf{h}^T\mathbf{d} + K_{21} a & K_{12}\mathbf{h}^T\mathbf{d} + K_{22} a   \\ \end{bmatrix}\begin{bmatrix} \mathbf{h}^T  &  a \\ \mathbf{d}^T\mathbf{S} + d_\mathrm{T}\mathbf{h}^T & \mathbf{d}^T\mathbf{h}+d_\mathrm{T} a    \\ \end{bmatrix}.
\end{split}
\end{equation}
Before continuing, let us note that we are only interested in the top left block because $\Delta\mathbf{S}_\mathrm{m} = \left[ \mathbf{\hat{W}}_\mathrm{m}^{-1}\right]_\mathcal{AA} - \mathbf{S}$. We obtain
\begin{equation}
    \Delta\mathbf{S}_\mathrm{m} = \left( K_{11}\mathbf{S}\mathbf{d}+K_{21}\mathbf{h}\right) \mathbf{h}^T + \left(K_{12}\mathbf{S}\mathbf{d} + K_{22}\mathbf{h}\right) \left( \mathbf{d}^T\mathbf{S} + d_\mathrm{T}\mathbf{h}^T\right).
\end{equation}
Let us now define $\mathbf{e} = \mathbf{S}\mathbf{d}$ and $\mathbf{f} = \mathbf{d}^T\mathbf{S}$. Since reciprocity imposes $\mathbf{S}=\mathbf{S}^T$, we have $\mathbf{e} = \mathbf{f}^T$.

\begin{equation}
    \Delta\mathbf{S}_\mathrm{m} =  \left( K_{11}+K_{12}d_\mathrm{T}\right)\mathbf{e}\mathbf{h}^T + \left(K_{21}+K_{22}d_\mathrm{T}\right)\mathbf{h} \mathbf{h}^T + K_{12}\mathbf{e}\mathbf{e}^T +K_{22}\mathbf{h} \mathbf{e}^T.
    \label{eq3_main}
\end{equation}

\clearpage
\section{Analysis of TRACK based on our system model}

The TRACK method proposed in Refs.~\cite{ma2014time,zhou2014focusing} can be revisited and understood based on the system model we use in the main text. As stated in the main text, TRACK postulates that $\mathbf{h}^\prime_\mathrm{TRACK}$ is collinear with $\mathbf{h}$, where
\begin{equation}
    \mathbf{h}^\prime_\mathrm{TRACK} = \Delta\mathbf{S}\mathbf{x}
    \label{track_definition}
\end{equation}
and $\mathbf{x}$ is an arbitrary non-zero wavefront.

\subsection{Target polarizability change}

Inserting $\mathbf{S}_\mathrm{p}=-k\mathbf{h}\mathbf{h}^T$ from Eq.~(2) in the main text into Eq.~(\ref{track_definition}) yields
\begin{equation}
    \mathbf{h}^\prime_\mathrm{TRACK,{p}} = \Delta\mathbf{S}\mathbf{x} = -k\mathbf{h}\mathbf{h}^T\mathbf{x} = -kz\mathbf{h},
    \label{track_p}
\end{equation}
where $z=\mathbf{h}^T\mathbf{x}$ is a complex-valued scalar. Given Eq.~(\ref{track_p}), it is obvious that $\mathbf{h}^\prime_\mathrm{TRACK,{p}}$ is collinear with $\mathbf{h}$.

A vulnerability of TRACK lies in the fact that, by chance, sometimes $\mathbf{x}$ will be (almost) orthogonal to $\mathbf{h}^\star$. In such a scenario, the probing wavefront perfectly avoids the target and therefore the measured output wavefront cannot be sensitive to the target's polarizability change. This explains why the average performance of TRACK falls slightly short of the optimal performance of SVD in Fig.~2 of the main text even in the high-SNR regime.

The assumption about unitarity made in Refs.~\cite{ma2014time,zhou2014focusing} does \textit{not} concern $\mathbf{S}$ but rather the transmission matrix $\mathbf{\Gamma}$ from the $N_\mathrm{A}$ antennas to a multitude of locations inside the medium, including that of the target. If $\mathbf{\Gamma}$ is approximately unitary, then focusing on the target with perfect \textit{contrast} is possible, i.e., the ratio of energy received by the target compared to the energy received at the other considered locations inside the medium tends to infinity. In practice, $\mathbf{\Gamma}$ is never perfectly unitary but for large values of $N_\mathrm{A}$ (in optical wavefront shaping experiments, $N_\mathrm{A}$ can reach and exceed the order of $10^3$), $\mathbf{\Gamma}^\dagger\mathbf{\Gamma}$ will almost approximate a scaled identity matrix if the medium is complex, implying close to perfect contrast. However, with respect to the definition of ``optimal focusing'' used in our Letter (we define ``optimal focusing'' as delivering as much power as possible to the target), the assumption that $\mathbf{\Gamma}$ is unitary or close to unitary does not play any role.

\subsection{Target motion}

Inserting $\mathbf{S}_\mathrm{m}= \left( K_{11}+K_{12}d_\mathrm{T}\right)\mathbf{e}\mathbf{h}^T + \left(K_{21}+K_{22}d_\mathrm{T}\right)\mathbf{h} \mathbf{h}^T + K_{12}\mathbf{e}\mathbf{e}^T +K_{22}\mathbf{h} \mathbf{e}^T$ from Eq.~(3) in the main text into Eq.~(\ref{track_definition}) yields
\begin{equation}
      \mathbf{h}^\prime_\mathrm{TRACK,{m}} = \Delta\mathbf{S}\mathbf{x} =\left( \left( K_{11}+K_{12}d_\mathrm{T}\right)\mathbf{e}\mathbf{h}^T + \left(K_{21}+K_{22}d_\mathrm{T}\right)\mathbf{h} \mathbf{h}^T + K_{12}\mathbf{e}\mathbf{e}^T +K_{22}\mathbf{h} \mathbf{e}^T \right) \mathbf{x}.
\end{equation}
Defining the complex-valued scalar $y = \mathbf{e}^T\mathbf{x}$, this simplifies to
\begin{equation}
\begin{split}
    \mathbf{h}^\prime_\mathrm{TRACK,{m}} = \left( K_{11}+K_{12}d_\mathrm{T}\right)z\mathbf{e} + \left(K_{21}+K_{22}d_\mathrm{T}\right)z\mathbf{h}  + K_{12}y\mathbf{e} +K_{22}y\mathbf{h} \\ = \left( \left( K_{11}+K_{12}d_\mathrm{T}\right)z + K_{12}y\right) \mathbf{e} +  \left( \left(K_{21}+K_{22}d_\mathrm{T}\right)z + K_{22} y \right) \mathbf{h} .
\end{split}
\label{track_m}
\end{equation}
Given Eq.~(\ref{track_m}), it is obvious that $ \mathbf{h}^\prime_\mathrm{TRACK,{m}}$ is a linear combination of $\mathbf{h}$ and $\mathbf{e}$ which is, in general, not collinear with $\mathbf{h}$.
While TRACK does hence not yield optimal focusing, TRACK nonetheless yields a much better focusing performance (averaged over frequencies) in Fig.~2 of the main text than RAND which disposes of no information about $\mathbf{h}$ other than that $\mathbf{h} \in \mathbb{C}^{N_\mathrm{A}}$. This is because TRACK enables the identification of a two-dimensional space that contains $\mathbf{h}$.

\clearpage
\section{Analysis of GWS based on our system model}

The GWS method proposed in Refs.~\cite{ambichl2017focusing,horodynski2020optimal} can be revisited and understood based on the system model we use in the main text. As stated in the main text, the starting point for the GWS method is
\begin{equation}
    \mathbf{Q}_\mathrm{q} = -\jmath \mathbf{S}^{-1} \Delta\mathbf{S}_\mathrm{q},
    \label{gws_definition}
\end{equation}
where we approximate $\partial \mathbf{S} / \partial q$ with $\Delta\mathbf{S}_\mathrm{q}$. $q$ denotes the perturbed parameter. 
If $\mathbf{S}$ is unitary, then $\mathbf{Q}_\mathrm{q}$ must be a Hermitian operator with real eigenvalues.
In the sub-unitary system we consider, the GWS method consists in defining $ \mathbf{h}^\prime_\mathrm{GWS,{q}} $ as the eigenvector of $\mathbf{Q}_\mathrm{q}$ associated with the eigenvalue that has the largest modulus.

\subsection{Target polarizability change}

Inserting $\Delta\mathbf{S}_\mathrm{p}=-k\mathbf{h}\mathbf{h}^T$ from Eq.~(2) in the main text into Eq.~(\ref{gws_definition}) yields
\begin{equation}
    \mathbf{Q}_\mathrm{p} = -\jmath \mathbf{S}^{-1} \Delta\mathbf{S}_\mathrm{p} = \jmath k \mathbf{S}^{-1} \mathbf{h}\mathbf{h}^T = \jmath k \mathbf{z}\mathbf{h}^T ,
    \label{gws_deltaS_p}
\end{equation}
where we have defined $\mathbf{z} = \mathbf{S}^{-1} \mathbf{h}$. Given Eq.~(\ref{gws_deltaS_p}), it is obvious that the \textit{conjugate} of the \textit{right} singular vector of $\mathbf{Q}_\mathrm{p}$ associated with the only non-zero singular value is collinear with $\mathbf{h}$. 
Moreover, the \textit{conjugate} of the \textit{left} eigenvector of $\mathbf{Q}_\mathrm{p}$ associated with the only non-zero eigenvalue is collinear with $\mathbf{h}$.  Hence, defining $ \mathbf{h}^\prime_\mathrm{GWS,{p}} $ as the conjugate of the left eigenvector of $\mathbf{Q}_\mathrm{p} $ associated with the largest (and only non-zero) eigenvalue yields a vector that is collinear with $\mathbf{h}$. 

A vulnerability of GWS seems to lie in the fact that, by chance, sometimes $\mathbf{S}$ will be (almost) rank-deficient in which case we heuristically observe that the approach tends to fail. 

The assumptions of (i) $\mathbf{S}$ being unitary and (ii) the perturbation being of infinitesimal size are necessary in Ref.~\cite{horodynski2020optimal} in order to justify the physical interpretation of the GWS operator. However, if the goal is simply to achieve optimal focusing, straightforward algebraic manipulations of our system model are sufficient to interpret the problem without a need to resort to the GWS operator and/or to make the strongly limiting foundational assumptions of the GWS approach.

\subsection{Target motion}

Inserting $\Delta\mathbf{S}_\mathrm{m}= \left( K_{11}+K_{12}d_\mathrm{T}\right)\mathbf{e}\mathbf{h}^T + \left(K_{21}+K_{22}d_\mathrm{T}\right)\mathbf{h} \mathbf{h}^T + K_{12}\mathbf{e}\mathbf{e}^T +K_{22}\mathbf{h} \mathbf{e}^T$ from Eq.~(3) in the main text into Eq.~(\ref{gws_definition}) yields
\begin{equation}
\begin{split}
    \mathbf{Q}_\mathrm{m} = -\jmath \mathbf{S}^{-1} \Delta\mathbf{S}_\mathrm{m}= -\jmath \mathbf{S}^{-1} \left( \left( K_{11}+K_{12}d_\mathrm{T}\right)\mathbf{e}\mathbf{h}^T + \left(K_{21}+K_{22}d_\mathrm{T}\right)\mathbf{h} \mathbf{h}^T + K_{12}\mathbf{e}\mathbf{e}^T +K_{22}\mathbf{h} \mathbf{e}^T \right) \\
 =    -\jmath \left( \left( K_{11}+K_{12}d_\mathrm{T}\right)\mathbf{d}\mathbf{h}^T + \left(K_{21}+K_{22}d_\mathrm{T}\right)\mathbf{z} \mathbf{h}^T + K_{12}\mathbf{d}\mathbf{e}^T +K_{22}\mathbf{z} \mathbf{e}^T   \right),
\end{split}
    \label{gws_motion}
\end{equation}
where we use $\mathbf{S}^{-1} \mathbf{e}=\mathbf{S}^{-1} \mathbf{S} \mathbf{d} = \mathbf{d} $. Given Eq.~(\ref{gws_motion}), it is obvious that the \textit{conjugates} of the first two \textit{right} singular vectors of $\mathbf{Q}_\mathrm{m}$ are linear combinations of $\mathbf{h}$ and $\mathbf{e}$, such that, in general, neither of them is collinear with $\mathbf{h}$. 
While GWS does hence not yield optimal focusing, GWS nonetheless yields a much better focusing performance (averaged over frequencies) in Fig.~2 of the main text than RAND which disposes of no information about $\mathbf{h}$ other than that $\mathbf{h} \in \mathbb{C}^{N_\mathrm{A}}$, as in the case of TRACK. This is because TRACK and GWS enable the identification of a two-dimensional space that contains $\mathbf{h}$.

\clearpage
\section{Analysis of polarizability-changing target in a dynamic complex medium}

In this supplementary note, we analyze the case of a polarizability-changing target inside a complex medium that also includes parasitic polarizability-changing perturbers.
In our experimental setup in Fig.~4 in the main text, the perturbers are emulated by randomly configured programmable meta-atoms~\cite{ahmed2023over}. The approach to model these perturbers as polarizability-changing dipoles was recently experimentally validated in Ref.~\cite{sol2023experimentally}.

In this case, our system comprises $N_\mathrm{A}$ antennas, one target, and $N_\mathrm{S}$ perturbers, yielding a total number of $N=N_\mathrm{A}+1+N_\mathrm{S}$ dipoles. Defining $\mathcal{S}$ as the set of dipole indices belonging to the parasitic perturbers, the interaction matrix at time instant $t$ can be partitioned into the following $3\times 3$ block form:
\begin{equation}
    \mathbf{\hat{W}}(t) = \begin{bmatrix} 
	\mathbf{\hat{W}}_\mathcal{AA}(t) & \mathbf{\hat{W}}_\mathcal{AT}(t) & \mathbf{\hat{W}}_\mathcal{AS}(t) \\ \mathbf{\hat{W}}_\mathcal{TA}(t) & 	\mathbf{\hat{W}}_\mathcal{TT}(t) & \mathbf{\hat{W}}_\mathcal{TS}(t) \\ \mathbf{\hat{W}}_\mathcal{SA}(t) & \mathbf{\hat{W}}_\mathcal{ST}(t) & \mathbf{\hat{W}}_\mathcal{SS}(t)\\  \end{bmatrix}.
\end{equation}
Assuming for concreteness (and ease of explicit writing) that $N_\mathrm{S}=3$, the inverse interaction matrix at time instant $t$ takes the following form:
\begin{equation}
    \left[\mathbf{\hat{W}}(t)\right]^{-1} = \begin{bmatrix} 
	\mathbf{S}(t) & \mathbf{h}(t) & \mathbf{l}_1(t) & \mathbf{l}_2(t) & \mathbf{l}_3(t) \\ \mathbf{h}^T(t) & 	a(t) & b_{\mathrm{T}1}(t) & b_{\mathrm{T}2}(t) & b_{\mathrm{T}3}(t) \\ \mathbf{l}_1^T(t) & b_{\mathrm{T}1}(t) & b_1(t) & b_{12}(t) & b_{13}(t)\\ \mathbf{l}_2^T(t) & b_{\mathrm{T}2}(t) & b_{12}(t) & b_{2}(t) & b_{23}(t)\\ \mathbf{l}_3^T(t) & b_{\mathrm{T}3}(t) & b_{13}(t) & b_{12}(t) & b_{3}(t)\\  \end{bmatrix},
\end{equation}
where $\mathbf{l}_v$ is the transmission vector from the antennas to the $v$th parasitic perturber.

Let us consider a very concrete realization for illustration purposes: the target, the first parasitic perturber and the third parasitic perturber change their polarizabilities between two measurements separated by a time interval $\Delta t$ whereas the second parasitic perturber does not change its polarizability. Then, according to the Woodbury identity, following Ref.~\cite{prod2023efficient} we have
\begin{equation}
    \mathbf{\hat{W}}^{-1} (t+\Delta t)-\mathbf{\hat{W}}^{-1} (t) = \mathbf{U}\mathbf{C}\mathbf{V},
    \label{woody_dynamic}
\end{equation}
where
\begin{subequations}
    \begin{equation}
        \mathbf{C} = \begin{bmatrix} \Delta\alpha_\mathrm{T}^{-1} & 0 & 0 \\ 0 & 	\Delta\alpha_\mathrm{P1}^{-1} & 0 \\ 0 & 0 & \Delta\alpha_\mathrm{P3}^{-1} \end{bmatrix},
    \end{equation}
    \begin{equation}
        \mathbf{U} = \mathbf{V}^T = \begin{bmatrix} \mathbf{0}_{N_\mathrm{A}} & 1 & 0 & 0 & 0 \\  \mathbf{0}_{N_\mathrm{A}} & 0 & 1 & 0 & 0 \\ \mathbf{0}_{N_\mathrm{A}} & 0 & 0 & 0 & 1 \end{bmatrix}^T,
    \end{equation}
\end{subequations}
where $\Delta\alpha_\mathrm{P1}^{-1}$ and $\Delta\alpha_\mathrm{P3}^{-1}$ are the inverse polarizability changes of the first and third parasitic perturbers. We further find that
    \begin{subequations}
 \begin{equation}
        \left[\mathbf{{\hat{W}}}(t)\right]^{-1} \mathbf{U} = \begin{bmatrix} 
	\mathbf{h}(t) & \mathbf{l}_1(t) & \mathbf{l}_3(t)  \\ 	a(t) & b_{\mathrm{T}1}(t) & b_{\mathrm{T}3}(t)  \\ b_{\mathrm{T}1}(t) & b_{1}(t) & b_{13}(t) \\ b_{\mathrm{T}2}(t) & b_{12}(t) & b_{23}(t) \\ b_{\mathrm{T}3}(t) & b_{13}(t) & b_{3}(t)\end{bmatrix}.
    \end{equation}
    \begin{equation}
         \mathbf{V} \left[\mathbf{{\hat{W}}}(t)\right]^{-1} \mathbf{U} = \left[\left[ \mathbf{\hat{W}}(t)\right]^{-1}\right]_\mathcal{TT} = \begin{bmatrix} 
	 	a(t) & b_{\mathrm{T}1}(t) &  b_{\mathrm{T}3}(t) \\ b_{\mathrm{T}1}(t) & b_1(t) &  b_{13}(t)\\  b_{\mathrm{T}3}(t) & b_{13}(t) & b_{3}(t)\\  \end{bmatrix}.
    \end{equation}
    \begin{equation}
        \mathbf{V} \left[\mathbf{{\hat{W}}}(t)\right] = \begin{bmatrix} 
	\mathbf{h}^T(t)  & 	a(t)  & b_{\mathrm{T}1}(t) & b_{\mathrm{T}2}(t) & b_{\mathrm{T}3}(t) \\ \mathbf{l}_1^T(t)  & 	b_{\mathrm{T}1}(t)   & b_{1}(t) & b_{12}(t) & b_{13}(t) \\ \mathbf{l}_3^T(t)  & 	b_{\mathrm{T}3}(t)   & b_{13}(t) & b_{23}(t) & b_{3}(t) \end{bmatrix}.
    \end{equation}        
    \end{subequations}
For ease of notation, we furthermore define
\begin{equation}
    \mathbf{J} = \begin{bmatrix} J_{11}  &  J_{12} &  J_{13} \\ J_{21} & J_{22} & J_{32}   \\ J_{31} & J_{32} & J_{33} \\ \end{bmatrix} = \left(\mathbf{C}^{-1} + \mathbf{V} \left[\mathbf{{\hat{W}}}(t)\right]^{-1} \mathbf{U} \right)^{-1}. 
\end{equation}
For the purposes of this discussion, it is not necessary to work out exact expressions for the entries of $\mathbf{J}$.
Substituting these results into Eq.~(\ref{woody_dynamic}), we obtain
\begin{equation}
\begin{split}
    \Delta \mathbf{S}(t) = \left[ \mathbf{\hat{W}}^{-1} (t+\Delta t)-\mathbf{\hat{W}}^{-1} (t) \right]_\mathcal{AA} \\ = J_{11}\mathbf{h}(t)\mathbf{h}^T(t) + J_{21}\mathbf{l}_1(t)\mathbf{h}^T(t) + J_{31}\mathbf{l}_3(t)\mathbf{h}^T(t) + J_{12}\mathbf{h}(t)\mathbf{l}_1^T(t)+ J_{22}\mathbf{l}_1(t)\mathbf{l}_1^T(t) \\+ J_{32}\mathbf{l}_3(t)\mathbf{l}_1^T(t) + J_{13}\mathbf{h}(t)\mathbf{l}_3^T(t) + J_{23}\mathbf{l}_1(t)\mathbf{l}_3^T(t) + J_{33}\mathbf{l}_3(t)\mathbf{l}_3^T(t)
    \end{split}.\label{eqs27sdfsd}
\end{equation}

By inspection of Eq.~(\ref{eqs27sdfsd}), it is clear that the left singular vectors of $\Delta\mathbf{S}(t)$ are linear combinations of $\mathbf{h}(t)$, $\mathbf{l}_1(t)$ and $\mathbf{l}_3(t)$. It makes sense that mathematically the transmission vectors to the target and to the parasitic perturbers play the same role in Eq.~(\ref{eqs27sdfsd}). While we considered a specific case to derive Eq.~(\ref{eqs27sdfsd}), it is obvious that in general the left singular vectors of $\Delta\mathbf{S}$ are linear combinations of the transmission vectors to all dipoles that changed their polarizability, as stated in the main text.

\clearpage
\section{Additional results for Fig.~4}

\begin{figure}[h]
    \centering
    \includegraphics[width=\columnwidth]{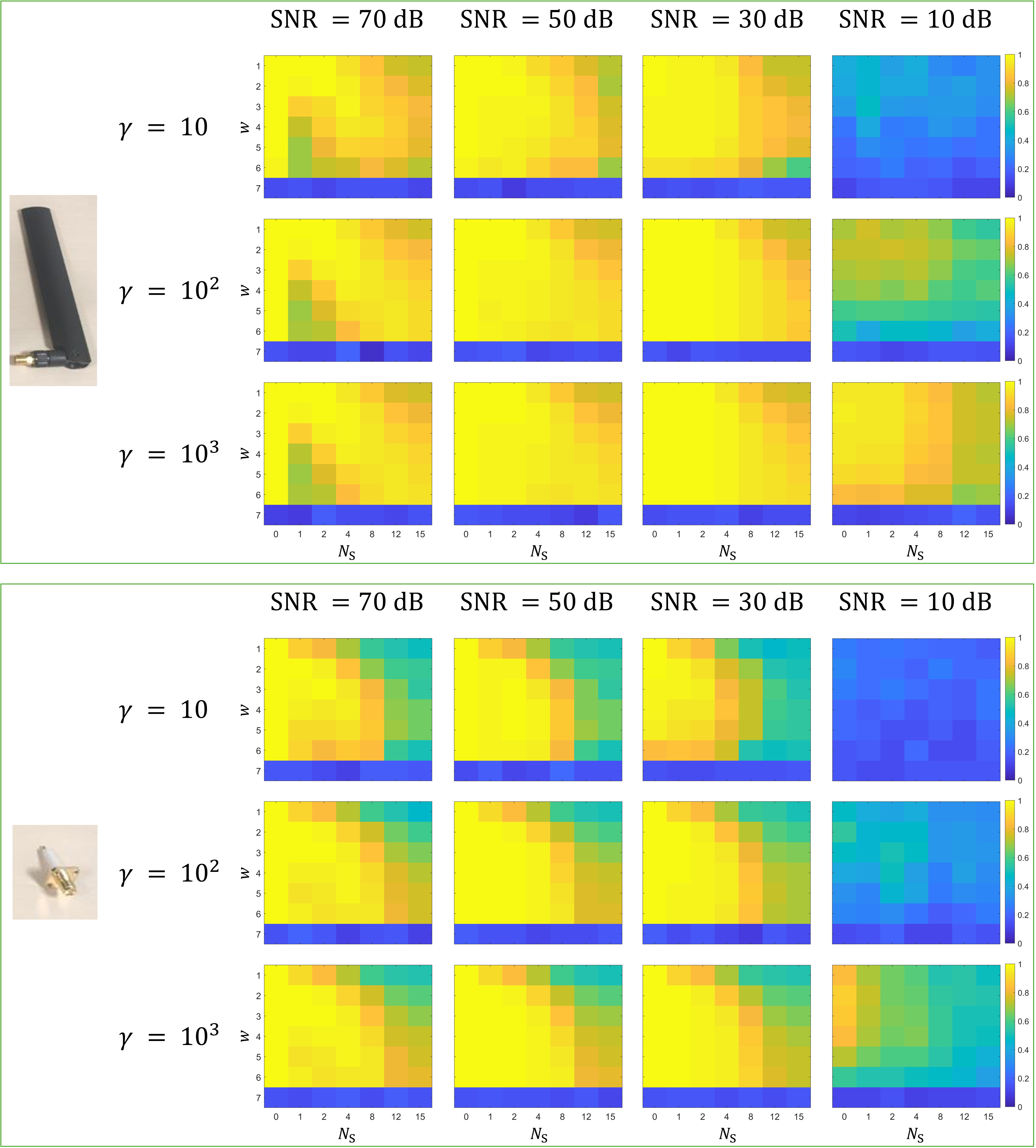}
    \caption{Systematic analysis of how $\gamma$, SNR and the target antenna type impact $\langle\tilde{\eta}_p\rangle_f$.}
    \label{FigS3}
\end{figure}

\clearpage
\section{Case of measuring $\mathbf{T}$ instead of $\mathbf{S}$}

In this supplementary note, we consider the case in which a subset $\mathcal{A_T}$ of the antennas can only transmit and the remaining subset $\mathcal{A_R} = \mathcal{A} \setminus \mathcal{A_T}$ of the antennas can only receive such that only an off-diagonal block $\mathbf{T}$ of $\mathbf{S}$ can be measured:
\begin{equation}
    \mathbf{S} = \begin{bmatrix} \mathbf{R^{in}} & \mathbf{T}^T\\ \mathbf{T} &  \mathbf{R^{out}}    \\ \end{bmatrix}.
\end{equation}
Consequently, $\mathbf{\hat{W}}$ can be partitioned as $3 \times 3 $ block matrix:
\begin{equation}
    \mathbf{\hat{W}} = \begin{bmatrix} \mathbf{S} & \mathbf{h}\\ \mathbf{h}^T & a    \\ \end{bmatrix} = \begin{bmatrix} \mathbf{R^{in}} & \mathbf{T}^T & \mathbf{t}\\ \mathbf{T} &  \mathbf{R^{out}} & \mathbf{r} \\ \mathbf{t}^T &  \mathbf{r}^T & a    \\ \end{bmatrix},
\end{equation}
where we use $\mathbf{h}^T = \left[ \mathbf{t}^T \ \mathbf{r}^T \right] $. $\mathbf{t}$ (resp. $\mathbf{r}$) is the transmission vector from the subset $\mathcal{A_T}$ (resp. $\mathcal{A_R}$) to the target.

\textit{Remark:} The case of being limited to measurements of a diagonal block of $\mathbf{S}$ such as $\mathbf{R^{in}}$ does not qualitatively differ from the case of being able to measure $\mathbf{S}$: each involved antenna transmits and receives, and the unobserved scattering through the remaining antennas is simply another contribution to the sub-unitarity of the considered matrix.

\subsection{Target polarizability change}

Substituting $\mathbf{h}^T = \left[ \mathbf{t}^T \ \mathbf{r}^T \right] $ into Eq.~(\ref{eq1main}) yields
\begin{equation}
  \mathbf{{\hat{W}}}_\mathrm{p}^{-1} = \mathbf{\hat{W}}^{-1} - \begin{bmatrix} 
	\mathbf{h}  \\ 	a  \\ \end{bmatrix} k  \begin{bmatrix} 
	\mathbf{h}^T  & 	a   \end{bmatrix} = \mathbf{\hat{W}}^{-1} - \begin{bmatrix} 
	\mathbf{t}  \\ \mathbf{r} \\ 	a  \\ \end{bmatrix} k  \begin{bmatrix} 
	\mathbf{t}^T  & \mathbf{r}^T  & 	a   \end{bmatrix} = \mathbf{\hat{W}}^{-1} - k \begin{bmatrix} 
	\mathbf{t}\mathbf{t}^T & \mathbf{t}\mathbf{r}^T & \mathbf{t}a  \\ \mathbf{r}\mathbf{t}^T & \mathbf{r}\mathbf{r}^T & \mathbf{r}a\\ 	a\mathbf{t}^T & a\mathbf{r}^T & a^2 \\ \end{bmatrix}.
\end{equation}
Hence, the measurement of $\Delta\mathbf{T}_\mathrm{p}$ must equal
\begin{equation}
    \Delta\mathbf{T}_\mathrm{p} = \left[ \mathbf{{\hat{W}}}_\mathrm{p}^{-1} \right]_\mathcal{A_R A_T} - \left[ \mathbf{{\hat{W}}}^{-1} \right]_\mathcal{A_R A_T} = -k \mathbf{r} \mathbf{t}^T,
\end{equation}
implying that $\Delta\mathbf{T}_\mathrm{p}$ is of rank one and the conjugate of its first right singular vector is collinear with $\mathbf{t}$. Hence we define $\mathbf{t}_\mathrm{p}^\prime$ to be equal to the conjugate of the first right singular vector of $\Delta\mathbf{T}_\mathrm{p}$ which allows us to identify the provably optimal wavefront for focusing on the target in the scenario considered in this supplementary note (only the antennas in $\mathcal{A_T}$ are allowed to transmit and only the antennas in $\mathcal{A_R}$ are allowed to receive).

\begin{figure}[h]
    \centering
    \includegraphics[width=0.8\columnwidth]{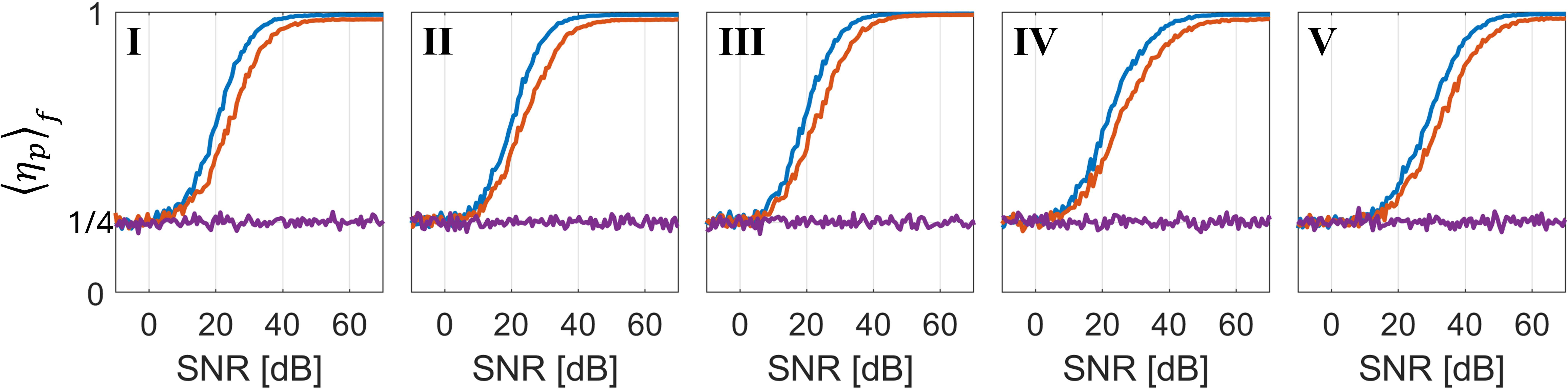}
    \caption{Focusing efficiency with SVD, GWS and RAND if only the off-diagonal block $\mathbf{T}$ of $\mathbf{S}$ can be measured, for the five target antenna types seen in Fig.~2 in the main text.}
    \label{FigS1}
\end{figure}

In Fig.~2 in the main text we analyzed as a function of the SNR the efficiency of focusing on a polarizability-changing target based on measurements of $\mathbf{S}$ for the proposed optimal SVD approach and the three benchmarks (TRACK, GWS, RAND). Here, we perform the same analysis if ony an off-diagonal block $\mathbf{T}$ of $\mathbf{S}$ is measured. In that case, the TRACK method cannot be applied (unless the transmitting and receiving antennas can switch their roles in which case one can find vectors collinear with $\mathbf{r}$ [resp.~$\mathbf{t}$] by transmitting a random wavefront with the antennas included in $\mathcal{T}$ [resp.~$\mathcal{R}$]). 
The GWS approach remains the same except that $\mathbf{T}^{-1}$ must be defined as the pseudo-inverse of $\mathbf{T}$ (since $\mathbf{T}$ is generally not a square matrix). We consider the same setting as in Fig.~2 with $\mathcal{A_T} = \{1,2,3,4\}$ and $\mathcal{A_R} = \{5,6,7\}$. The results in Fig.~\ref{FigS1} confirm that the proposed SVD approach achieves optimal focusing in the high-SNR regime and slightly outperforms the GWS approach, analogous to the results in Fig.~2.

\subsection{Target motion}

In addition to $\mathbf{h}^T = \left[ \mathbf{t}^T \ \mathbf{r}^T \right] $, we now define $\mathbf{e}^T = \left[ \mathbf{e_t}^T \ \mathbf{e_r}^T \right] $ and insert these into Eq.~(\ref{eq3_main}):
\begin{equation}
\begin{split}
    \Delta\mathbf{S}_\mathrm{m} =  \left( K_{11}+K_{12}d_\mathrm{T}\right)\begin{bmatrix} 
	\mathbf{e_t} \\ \mathbf{e_r}  \end{bmatrix}\begin{bmatrix} 
	\mathbf{t}^T & \mathbf{r}^T  \end{bmatrix} + \left(K_{21}+K_{22}d_\mathrm{T}\right)\begin{bmatrix} 
	\mathbf{t} \\ \mathbf{r}  \end{bmatrix} \begin{bmatrix} 
	\mathbf{t}^T & \mathbf{r}^T  \end{bmatrix} + K_{12}\begin{bmatrix} 
	\mathbf{e_t} \\ \mathbf{e_r}  \end{bmatrix}\begin{bmatrix} 
	\mathbf{e_t}^T & \mathbf{e_r}^T  \end{bmatrix} +K_{22}\begin{bmatrix} 
	\mathbf{t} \\ \mathbf{r}  \end{bmatrix} \begin{bmatrix} 
	\mathbf{e_t}^T & \mathbf{e_r}^T  \end{bmatrix}.
    \label{eq3_main_T}
\end{split}
\end{equation}
Then, the bottom left block reads
\begin{equation}
    \Delta\mathbf{T}_\mathrm{m} = \left[ \Delta\mathbf{S}_\mathrm{m} \right]_\mathcal{A_R A_T} = \left( K_{11}+K_{12}d_\mathrm{T}\right) \mathbf{e_r} \mathbf{t}^T  + \left(K_{21}+K_{22}d_\mathrm{T}\right) \mathbf{r}\mathbf{t}^T + K_{12}\mathbf{e_r}\mathbf{e_t}^T + K_{22}\mathbf{r}\mathbf{e_t}^T.
\end{equation}

Analogous to the discussion for the search of $\mathbf{h}$ based on $\Delta\mathbf{S}_\mathrm{m}$ in the main text, we observe here for the search of $\mathbf{t}$ based on $ \Delta\mathbf{T}_\mathrm{m}$ that $ \Delta\mathbf{T}_\mathrm{m}$ is a matrix of rank two whose first two right singular vectors span a space defined by the conjugates of $\mathbf{t}$ and $\mathbf{e_t}$ such that the first right singular vector is a linear combination of the conjugates of $\mathbf{t}$ and $\mathbf{e_t}$  and hence, in general, \textit{not} collinear with $\mathbf{t}$ or its conjugate. However, given three measurements $\mathbf{T}_\mathrm{A}$, $\mathbf{T}_\mathrm{B}$ and $\mathbf{T}_\mathrm{C}$ corresponding to three arbitrary distinct positions A, B and C of the target, we can evaluate the two rank-two matrices $\Delta\mathbf{T}_\mathrm{m}^\mathrm{AC} = \mathbf{T}_\mathrm{C}-\mathbf{T}_\mathrm{A}$ and $\Delta\mathbf{T}_\mathrm{m}^\mathrm{BC}= \mathbf{T}_\mathrm{C}-\mathbf{T}_\mathrm{B}$ (assuming, without loss of generality, that we wish to focus on the target while it is at position C). Then, we extract $\mathbf{t_m^\prime}$ as the intersection of the spaces spanned by the conjugates of the two first right singular vectors of $\Delta\mathbf{T}_\mathrm{m}^\mathrm{AC}$ and $\Delta\mathbf{T}_\mathrm{m}^\mathrm{BC}$ with the same approach as in the main text for $\mathbf{h}_\mathrm{m}^\prime$. $\mathbf{t}_\mathrm{m}^\prime$ is collinear with $\mathbf{t}$ and hence provably optimal for focusing on the target while it is at position C in the supplementary note considered in this section (only the antennas in $\mathcal{A_T}$ are allowed to transmit and only the antennas in $\mathcal{A_R}$ are allowed to receive).

\begin{figure}[h]
    \centering
    \includegraphics[width=0.7\columnwidth]{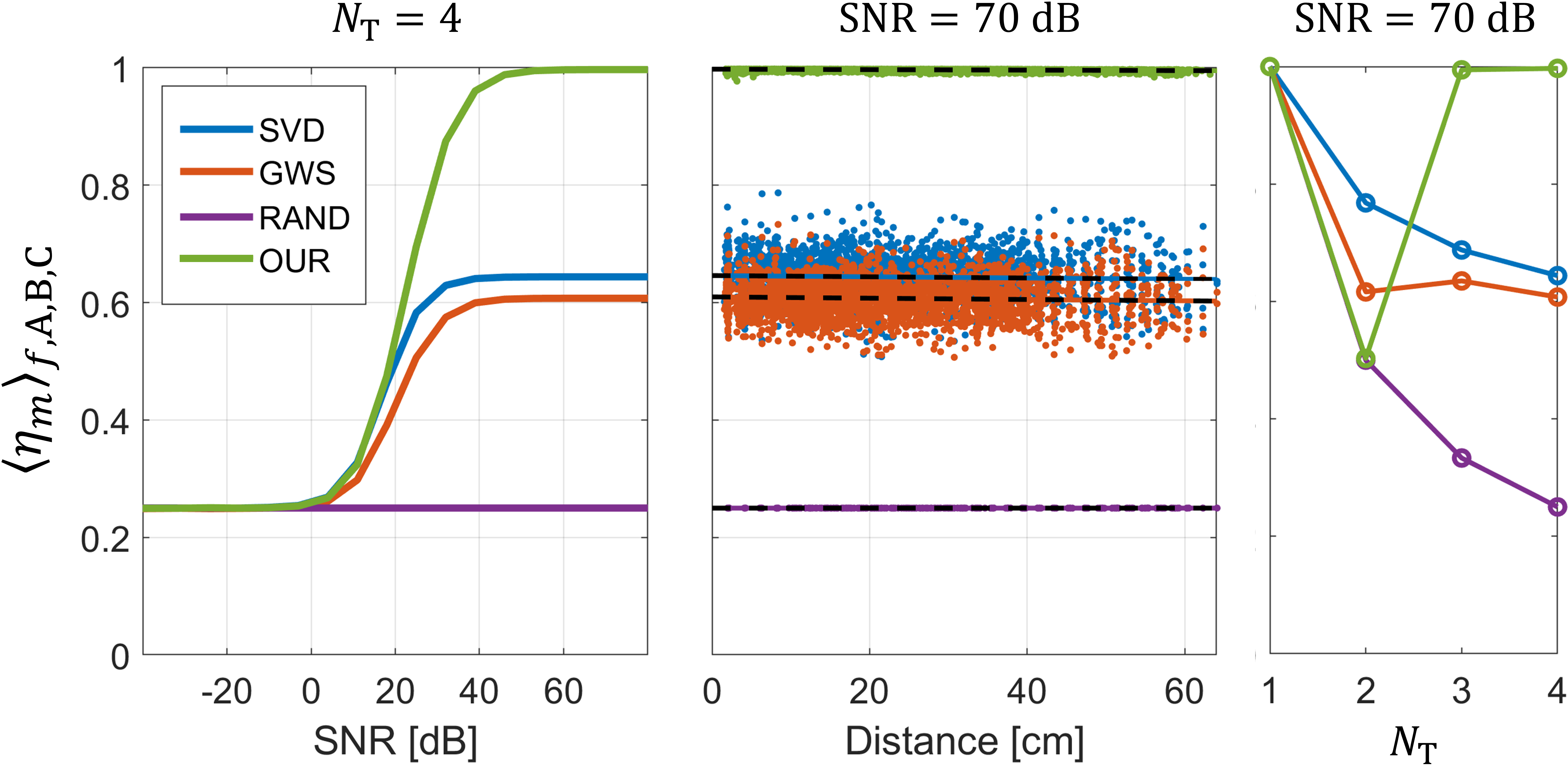}
    \caption{Focusing efficiency with SVD, GWS, RAND and OUR approaches if only the off-diagonal block $\mathbf{T}$ of $\mathbf{S}$ can be measured, as a function of the SNR, the distance between the positions A, B and C, and the number of transmitting antennas $N_\mathrm{T}$.}
    \label{FigS2}
\end{figure}

We consider the same setting as in Fig.~2 but with $\mathcal{A_T} = \{1,2,3,4\}$ and $\mathcal{A_R} = \{5,6,7\}$. As for Fig.~\ref{FigS1}, the TRACK method cannot be applied and the GWS method requires an approximation of $\mathbf{T}^{-1}$ by the pseudo-inverse of $\mathbf{T}$. 
The results in Fig.~\ref{FigS2} echo those from Fig.~3 in the main text: for $N_\mathrm{T}>2$, OUR approach achieves optimal focusing at sufficiently high SNR levels, irrespective of the distances between A, B and C. In contrast, the performance of SVD and the literature benchmark GWS remains substantially below optimal performance and deteriorates as $N_\mathrm{T}$ increases; SVD and GWS also slightly deteriorate as a function of distance. The dependence of the focusing efficiency on $N_\mathrm
T$ shown in Fig.~\ref{FigS2} assumes that only the first $N_\mathrm{T}$ antennas included in $\mathcal{A_T} = \{1,2,3,4\}$ are used as transmitters while always use all three receivers from $\mathcal{A_R} = \{5,6,7\}$ are used.
Recall that for $N_\mathrm{T}=1$, the input wavefront is a scalar and hence any random wavefront achieves $100\%$ focusing efficiency by definition. 
For $N_\mathrm{T}=2$, OUR approach cannot guarantee $100\%$ focusing efficiency because the stacked matrix is composed of linear combinations of the conjugates of three vectors (one of which is $\mathbf{h}$) but only has two non-zero singular values due to its dimensions. Specifically, for $N_\mathrm{T}=2$, both the ambient space $\mathbb{C}^2$ as well as the spans of $\Delta\mathbf{T}_\mathrm{m}^\mathrm{AC}$ and $\Delta\mathbf{T}_\mathrm{m}^\mathrm{BC}$ are two-dimensional and hence equal: $ \mathbb{C}^2 = \text{span}(\Delta\mathbf{T}_\mathrm{m}^\mathrm{AC}) = \text{span}(\Delta\mathbf{T}_\mathrm{m}^\mathrm{BC})$. Finding the intersection of the spans of $\Delta\mathbf{T}_\mathrm{m}^\mathrm{AC}$ and $\Delta\mathbf{T}_\mathrm{m}^\mathrm{BC}$ thus amounts to choosing a random vector in $\mathbb{C}^2$; indeed, OUR method is equivalent to RAND for $N_\mathrm{T}=2$, as seen in Fig.~\ref{FigS2}.

\clearpage
\section{Experimental details}

As seen in Fig.~2 and Fig.~4 in the main text, our complex medium is an irregularly shaped metallic enclosure of dimensions $59\times 60 \times 58 \mathrm{cm}^3$. Note that its top wall is removed to show its interior in Fig.~2 and Fig.~4. The holes seen in Fig.~3 for the experiment on target displacement are drilled into the top wall.

Seven identical antennas (ANT-24G-HL90-SMA; identical to target antenna type IV) were coupled to the system in order to input and output waves. As seen in Fig.~2 and Fig.~4, the seven antennas were regularly spaced, although their spatial arrangement has no influence on the presented results given the chaoticity of the system. 
Based on the average impulse response envelope's decay time, the composite quality factor of our system is found to be $Q=758$, implying a modal overlap of $\mathcal{N}=4$. 

The target antenna in Fig.~3 is a metallic screw that is electrically connected to the inner conductor of an SMA connector. No electric contact between the screw and the outer conductor exists.

All measurements are made with a 8-port vector network analyzer (Keysight M9005A) with 201 frequency points linearly spaced between 2~GHz and 3~GHz. The intermediate-frequency bandwidth is 1~kHz and the emitted power is 13~dB.
The cables connecting the VNA to the seven antennas are included in the calibration such that the calibration plane is where the antennas are connected to the cables. The cable connected to the target antenna in Fig.~2 and Fig.~4 is composed of two segments, one inside the cavity and one outside the cavity. Only the outside segment is included in the calibration such that the load impedance of the antenna can be changed without opening the cavity. This implies that the cable segment inside the cavity is effectively part of the target antenna in that case. The cable connected to the target antenna in Fig.~3 is calibrated up to the plane where it is connected to the antenna.

$\lVert \mathbf{S}^\dagger \mathbf{S} \rVert_2$ varies roughly between 0.3 and 0.7 in our experiments, evidencing that our system is clearly sub-unitarity. Most of the energy decay is of irreversible nature, originating from homogeneous absorption due to Ohmic losses on the cavity walls.

The 1-bit-programmable metasurface design is the same as in Ref.~\cite{ahmed2023over}. The operating bandwidth lies in the range $2.39\ \mathrm{GHz} - 2.51\ \mathrm{GHz}$, as seen in Fig.~3 of the main text. The configuration is digitally imposed with an Arduino microcontroller.

\end{document}